1# Optical attenuation without absorption

I.C. Nodurft, R.A. Brewster, T.B. Pittman and J.D. Franson

*Physics Department, University of Maryland Baltimore County, Baltimore, MD 21250, USA*
We consider a coherent state of light propagating through an ensemble of two-level atoms where all the atoms are initially in their ground state. In ordinary absorption, the transition of atoms to their excited state along with the absorption of a photon will remove energy from the beam and attenuate the signal. Here we show that post-selecting on those cases in which none of the atoms made a transition to the excited state can give even more attenuation than would normally occur due to absorption. The same process can also produce amplification when there is a sufficiently strong interaction between the photons and the atoms.
I. Introduction

Post-selection and heralding are used in quantum optics for a wide range of applications, such as linear optics quantum logic gates [1-5], noiseless amplification [6-9] and attenuation [10-12], and the quantum engineering of non-classical states [13-15]. Here we consider the effects of post-selection when a coherent state of light (laser beam) propagates through an atomic medium where all of the atoms are initially in their ground state. Without any post-selection, the absorption of photons and the corresponding transition of atoms to their excited state will remove energy from the beam and attenuate the signal.

We consider the effects of post-selecting on the case in which all of the atoms are found in their ground state at the end of the process as illustrated in Fig. 1. Since none of the atoms made a transition, one might expect that the atomic medium did nothing and that there should be no effect on the statistics of the photon state. Somewhat surprisingly, we show that post-selection of this kind can produce more attenuation than normally occurs due to absorption. It can even produce amplification under conditions where atomic saturation and time-dependent effects become important.

This work was motivated in part by an earlier paper [12] that showed that an optical parametric amplifier (OPA) can function as a noiseless attenuator if no photons are found in the output of the idler mode. Since the idler mode initially contained no photons, it can be inferred that the OPA did not emit or absorb any signal photons, since the signal and idler photons are emitted or absorbed in pairs. Once again, the OPA appears to have done nothing under these conditions, but the post-selection process gives noiseless attenuation nevertheless.

These effects can be understood from the fact that an incident coherent state contains an uncertain number of photons. Those probability amplitudes in the initial state that correspond to a relatively large number of incident photons are more likely to produce a transition to the excited atomic state and be rejected by the post-selection process. As a result, the photon number distribution is shifted towards lower values as illustrated in Fig. 2. Energy is conserved, even though the expectation value of the photon number has changed with no change in the energy of the environment.

The probability of success for the post-selection process decreases exponentially for high-intensity coherent states, where it becomes increasingly unlikely that no atoms will have been excited. As a result, this technique is limited to relatively weak input signals.

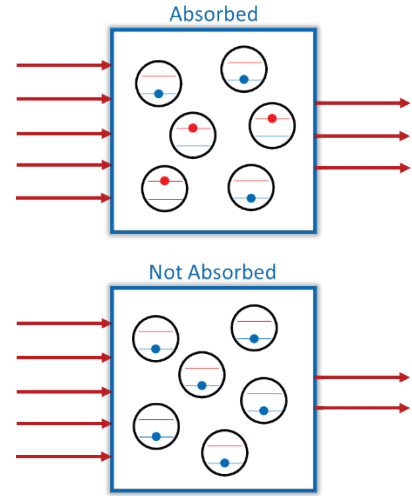

**Figure 1:** A coherent state of light is incident on an ensemble of two-level atoms in which all of the atoms are initially in their ground state (blue line). (a) Typical absorption in which some atoms are left in their excited state (red line) due to the absorption of a photon, which produces an attenuation of the beam of light. (b) Post-selection on the case in which no atoms are found in their excited state produces even more attenuation than does ordinary absorption.

We describe our analysis methods in Section II, where the density matrix of the system of atoms and photons is calculated with or without any post-selection. Section III presents the results when the interaction is relatively weak, giving enhanced attenuation in the absence of any absorption. Section IV considers the effects of relatively strong interactions which can produce either attenuation or amplification. A summary and conclusions are presented in Section V.

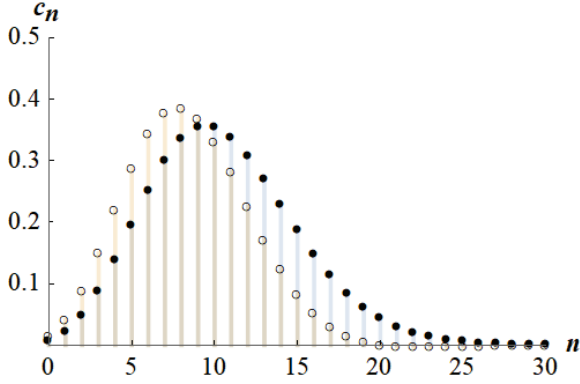

**Figure 2:** Origin of the increased attenuation due to the post-selection process shown in Fig. 1. The probability amplitude $c_n$ in the number state basis is plotted as a function of the photon number $n$ for a relatively weak interaction. The solid dots correspond to an incident coherent state while the open dots represent the state after the post-selection process. All of the probability amplitudes are initially reduced, but those corresponding to larger values of $n$ are reduced more than those corresponding to smaller values of $n$. This is because the interaction is stronger for large $n$ and the atoms are less likely to remain in the ground state and survive the post-selection process. The final state has been renormalized, which is why the probability amplitudes for small values of $n$ are larger than they were before the interaction.

## II. Analysis methods

We first consider the effects of post-selection when a coherent state interacts with a single atom, which illustrates most of the phenomena of interest. The results will then be generalized to an interaction with a larger number of atoms.

In the rotating wave approximation, the interaction between the photons and a single atom is described by the Jaynes-Cummings Hamiltonian, which is given by

$$\hat{H}_I = \hbar\lambda(\hat{a}^\dagger \hat{\sigma}_- + \hat{a}\hat{\sigma}_+) \qquad (1)$$

in the interaction picture [16,17]. Here $\lambda$ is the coupling constant between the light and an atom, which is proportional to the atomic dipole moment, and the operators $\hat{a}$ and $\hat{a}^\dagger$ annihilate or create a photon. The Pauli operator $\hat{\sigma}_+$ produces an atomic transition from the ground to excited state, while $\hat{\sigma}_-$ does the reverse operation. The Hamiltonian of Eq. (1) gives rise to a time evolution operator $\hat{U}(t)$ given as usual by

$$\hat{U}(t) = e^{-i\hat{H}_I t/\hbar} = e^{-i\lambda t(\hat{a}^\dagger \hat{\sigma}_- + \hat{a}\hat{\sigma}_+)}, \qquad (2)$$

where we have assumed that the interaction occurs over a time interval $t$.

Eq. (2) can be rewritten in block matrix form as [16]

$$\hat{U}(t) = \begin{bmatrix} \hat{C}(t) & \hat{S}'(t) \\ \hat{S}(t) & \hat{C}'(t) \end{bmatrix}. \qquad (3)$$

Here we have chosen a basis for the atomic states $|\psi_A\rangle$ in which

$$|\psi_A\rangle = \begin{pmatrix} c_E \\ c_G \end{pmatrix}, \qquad (4)$$

where $c_G$ and $c_E$ are the probability amplitudes for the ground and excited states, respectively. The elements of the block matrix in Eq. (3) are then given by

$$\hat{C}(t) = \cos\left(\lambda t \sqrt{\hat{a}\hat{a}^\dagger}\right) \qquad (5)$$

$$\hat{S}(t) = -i\hat{a}^\dagger \frac{\sin\left(\lambda t \sqrt{\hat{a}\hat{a}^\dagger}\right)}{\sqrt{\hat{a}\hat{a}^\dagger}} \qquad (6)$$

$$\hat{C}'(t) = \cos\left(\lambda t \sqrt{\hat{a}^\dagger \hat{a}}\right) \qquad (7)$$

$$\hat{S}'(t) = -i\hat{a}\frac{\sin\left(\lambda t \sqrt{\hat{a}^\dagger \hat{a}}\right)}{\sqrt{\hat{a}^\dagger \hat{a}}}. \qquad (8)$$

Before the interaction begins, we assume a coherent state $|\alpha\rangle$ for the field and we also assume that the atom is in its ground state. Since this is a pure state, the initial density operator describing the system is given by

$$\hat{\rho}(0) = \hat{\rho}^p(0) \otimes \hat{\rho}^A(0) = \begin{bmatrix} 0 & 0 \\ 0 & \hat{\rho}^p(0) \end{bmatrix}. \qquad (9)$$

Here $\hat{\rho}^p(0) = |\alpha\rangle\langle\alpha|$ and $\hat{\rho}^A(0) = |G\rangle\langle G|$ are the initial density operators for the electromagnetic field and the atom respectively. The interaction between

the field and the atom results in a time-evolved density operator $\hat{\rho}(t) = \hat{U}(t)\hat{\rho}(0)\hat{U}^\dagger(t)$. The new density operator is thus

$$\hat{\rho}(t) = \begin{bmatrix} \hat{\rho}_{11}(t) & \hat{\rho}_{12}(t) \\ \hat{\rho}_{21}(t) & \hat{\rho}_{22}(t) \end{bmatrix}, \quad (10)$$

with block matrix elements given by

$$\hat{\rho}_{11}(t) = -\hat{S}'(t)\hat{\rho}^p(0)\hat{S}(t) \quad (11)$$
$$\hat{\rho}_{12}(t) = \hat{S}'(t)\hat{\rho}^p(0)\hat{C}'(t) \quad (12)$$
$$\hat{\rho}_{21}(t) = -\hat{C}'(t)\hat{\rho}^p(0)\hat{S}(t) \quad (13)$$
$$\hat{\rho}_{22}(t) = \hat{C}'(t)\hat{\rho}^p(0)\hat{C}'(t). \quad (14)$$

We now consider two separate situations in which we either apply post-selection based on the final state of the atom, or we consider ordinary absorption in which the final state of the atom is ignored. For the case of ordinary absorption, we average over the atomic states by taking a partial trace. For a bipartite system of this kind, the reduced density matrix elements $\rho'_{nm}$ for the field after a partial trace over the atomic states is given by

$$\rho'_{nm} = \sum_\mu \rho_{n\mu,m\mu}. \quad (15)$$

Here the index $\mu$ labels the two-level atomic Hilbert space. The partial trace of Eq. (15) applied to the density matrix of Eq. (10) gives

$$\hat{\rho}'_p(t) = \hat{\rho}_{11}(t) + \hat{\rho}_{22}(t), \quad (16)$$

which describes the results of ordinary absorption by the atomic medium.

If, instead, we post-select on the case in which the atom is found in its ground state after the interaction, the new density matrix is found by projecting the original density operator onto the atomic ground state using $|G\rangle\langle G|\hat{\rho}(t)|G\rangle\langle G|$ and then normalizing. From Eq. (10), the resulting density operator for the field alone is given by

$$\hat{\rho}''_p(t) = \frac{\hat{\rho}_{22}(t)}{Tr[\hat{\rho}_{22}(t)]}. \quad (17)$$

Since all the terms in Eqs. (16) and (17) include only $\hat{\rho}_{11}(t)$ and $\hat{\rho}_{22}(t)$, we can ignore the off-diagonal terms. Using a Fock (number) state basis to describe the photons, it can be shown that the terms of interest are given by

$$\hat{\rho}_{11}(t) = e^{-|\alpha|^2} \sum_{n,m} c_{n,m} |n-1\rangle\langle m-1|$$
$$\hat{\rho}_{22}(t) = e^{-|\alpha|^2} \sum_{n,m} c'_{n,m} |n\rangle\langle m|, \quad (18)$$

where

$$c_{n,m} = \frac{\alpha^n \alpha^{*m}}{\sqrt{n!m!}} \sin(\lambda t \sqrt{n}) \sin(\lambda t \sqrt{m})$$
$$c'_{n,m} = \frac{\alpha^n \alpha^{*m}}{\sqrt{n!m!}} \cos(\lambda t \sqrt{n}) \cos(\lambda t \sqrt{m}). \quad (19)$$

Inserting Eqs. (18) and (19) into Eqs. (16) and (17) gives the explicit form of the final density operators for the two cases of interest. In the case of an interaction with $N$ atoms, the constants $c_{n,m}$ and $c'_{n,m}$ simply takes the pairs of sine or cosine functions to the $N$th power. These can be used to calculate the expectation value of the properties of the system, such as the mean number $\langle \hat{n} \rangle$ of photons given by

$$\langle \hat{n} \rangle = Tr[\hat{n}\hat{\rho}]. \quad (20)$$

Here $\hat{n} = \hat{a}^\dagger \hat{a}$ is the number operator. The mean photon number can be shown to be

$$\langle \hat{n} \rangle_1 = |\alpha|^2 - e^{-|\alpha|^2} \sum_{n=0}^{\infty} \frac{|\alpha|^{2n}}{n!} \sin^2(\lambda t \sqrt{n}) \quad (21)$$

for normal absorption, and

$$\langle \hat{n} \rangle_2 = |\alpha|^2 \frac{\sum_{n=0}^{\infty} \frac{|\alpha|^{2n}}{n!} \cos^2(\lambda t \sqrt{n+1})}{\sum_{n=0}^{\infty} \frac{|\alpha|^{2n}}{n!} \cos^2(\lambda t \sqrt{n})} \quad (22)$$

for post-selection on no absorption. The mean photon number will be used to quantify the amount of attenuation or amplification in the next two sections. The final density operators can also be used to calculate a quasi-probability distribution in phase-space for the final field, as will be described in the following sections.

The analytic results in Eqs. (21) and (22) were verified numerically by calculating the time evolution of the density matrix using Mathematica. The effects of an interaction with a larger number of atoms was also calculated numerically. For simplicity, we assumed that the photons interacted with a series of $N$ atoms one at a time. This illustrates all of the features of interest, and a situation of this kind could be realized experimentally by sending a narrow beam of light through an atomic vapor with a sufficiently low density that only a single atom passes through the beam at any given time.

In order to calculate the results of such a sequence of interactions, the photons were assumed to interact with the first atom as described by Eqs. (9) through (10). The reduced density matrix of the field was then calculated by tracing over the atomic states for ordinary absorption, or by projecting onto the subspace corresponding to the ground state of the atom for post-selection. The tensor product with the ground state of the next atom was formed and the process was repeated $N$ times. The results were qualitatively similar to those from an interaction with a single atom, except that the change in the state of the field was much larger as would be expected. The results of these calculations are discussed in the next two sections for an arbitrary choice of $N = 10$.

A Taylor series expansion of Eqs. (21) and (22) can be used to show that the post-selection and normal absorption processes give the same amount of attenuation in the limit of weak interactions or small coherent state amplitudes, as will be evident in the examples considered in the following sections.

The probability of success of the post-selection process can be calculated by taking the trace of the unnormalized density operators in Eq. (18). For an interaction with a single photon, the probability of success can be shown to be

$$P_1(t) = e^{-|\alpha|^2} \sum_{n=0}^{\infty} \frac{\alpha^n}{n!} \cos^2\left(\lambda t \sqrt{n}\right). \quad (23)$$

Extending this to an interaction with $N$ atoms gives a success probability of

$$P_N(t) = e^{-|\alpha|^2} \sum_{n=0}^{\infty} \frac{\alpha^n}{n!} \cos^{2N}\left(\lambda t \sqrt{n}\right). \quad (24)$$

It can be seen that the probability of success decreases exponentially for large values of $|\alpha|$, and it is on the order of $10^{-3}$ for most of the examples discussed in the following sections.

### III. Weak interactions and enhanced attenuation

In this section, we will consider the case in which the interaction between the field and the atoms is sufficiently weak that there is negligible saturation of the excited atomic states. Post-selecting on those events in which the atoms remained in their ground state gives enhanced attenuation in that case. In the following section, we will consider the more complicated situation where the interaction is sufficiently strong that atomic saturation and Rabi oscillations can play an important role.

The strength of the interaction between the field and an atom is characterized by the parameter $r = \lambda t$, which appears in the Hamiltonian of Eq (1) and all of the subsequent results. For simplicity, we will refer to $r$ as the interaction strength or coupling parameter.

The mean number of photons $\langle \hat{n} \rangle$ left in the field after an interaction with a single atom is illustrated in Fig. 3 as a function of the parameter $r$ over a relatively small range of values up to $r = 0.4$. For simplicity, the fraction of the photons remaining is plotted instead of $\langle \hat{n} \rangle$ itself. The initial amplitude of the coherent state was arbitrarily chosen to be $\alpha = \sqrt{10}$, which corresponds to a mean photon number of $10$. It can be seen that the post-selection process corresponding to no atomic absorption can give significantly more attenuation than is obtained from the usual absorption process for values of $r$ greater than approximately 0.2. The post-selected case and the usual absorption process give equivalent absorption in the limit of small $r$, as can be shown to be the case analytically.

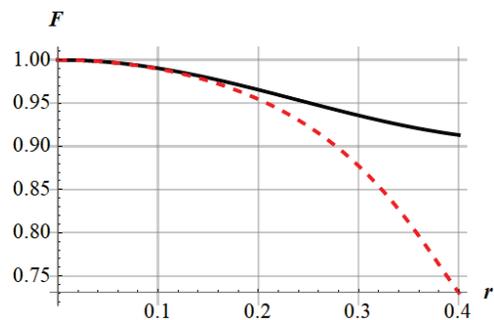

**Figure 3:** A plot of the fraction $F$ of the photons remaining on average after a coherent state interacts with a single atom. The black (solid) curve corresponds to ordinary absorption without post-selection, while the results based on post-selection on the atomic ground state are shown by the red (dashed) curve. It can be seen that post-selection on those cases where no atomic absorption occurred can give more attenuation than the usual atomic absorption process.

The effects of the post-selection can be further illustrated by plotting the Husimi-Kano Q-function in phase space, which is defined as [18,19]

$$Q(\alpha) = \frac{1}{\pi} \langle \alpha | \hat{\rho} | \alpha \rangle. \quad (25)$$

Here $\alpha$ is an arbitrary complex variable and $|\alpha\rangle$ is a coherent state with that amplitude. Fig. 4(a) shows the Q-function for the case of ordinary absorption with no post-selection, while Fig. 4(b) shows the results when post-selected on the case when all of the atoms remained in the ground state. These results correspond to $N=10$ and $r=0.25$, and were calculated numerically as described in the preceding section. It can be seen that the Q-function is shifted closer to the origin for the post-selected case in Fig. 4(b), which corresponds to a lower intensity than is the case with no post-selection in Fig. 4(a). The mean photon number corresponds to $\langle \hat{n} \rangle = 5.88$ for normal absorption and $\langle \hat{n} \rangle = 4.92$ for post-selection. It can be seen that the post-selection process also produces a slight distortion in the shape of the Q-function, which means that the process is somewhat nonlinear.

The probability of success for the post-selection process can be calculated from Eqs. (23) and (24). The probability of success under the conditions corresponding to Fig. 4 is on the order of $10^{-3}$, while it becomes exponentially smaller for larger coherent state amplitudes.

### IV. Strong interactions and amplification

Atomic saturation can become important when the coupling parameter $r$ is sufficiently large, and the atoms can undergo Rabi oscillations as well for large values of $r$. This results in a more complicated response of the field in which the system oscillates between loss and gain, as can be seen in Fig. 5.

Fig. 5 shows the fraction $F$ of the photons remaining as a function of $r$ as in Fig. 3, but for larger values of $r$ up to 3.0. It can be seen that the number of photons remaining at the end of the interaction now oscillates as a function of $r$, with certain values of $r$ producing an increase in the mean photon number rather than a decrease. This corresponds to a new kind of gain mechanism that occurs in the absence of any atomic transitions.

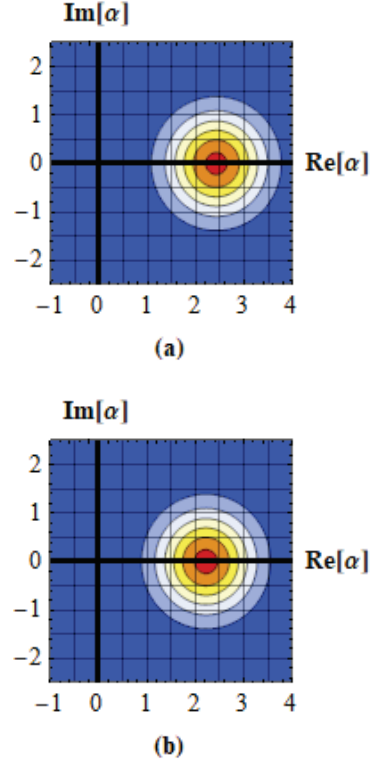

**Figure 4:** A plot of the Q-function of the field after an interaction with $N=10$ atoms and a coupling parameter of $r=0.25$. (a) The Q-function for ordinary atomic absorption where there is no post-selection. (b) The corresponding plot for post-selection on those events in which none of the atoms made a transition to the excited state. It can be seen that the post-selection process gives more attenuation than the usual atomic absorption.

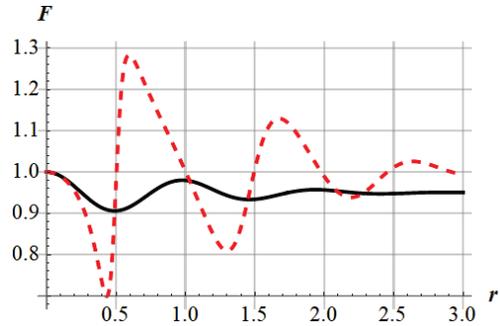

**Figure 5:** A plot of the fraction $F$ of the photons remaining on average as a function of the coupling parameter $r$ as in Fig. 3, but plotted over a range of $r$ where atomic saturation and Rabi oscillations become important. Once again, the black (solid) curve corresponds to ordinary absorption without post-selection, while the results based on post-selection on the atomic ground state are shown by the red (dashed) curve. It can be seen that the effect on the field now oscillates between attenuation and amplification.

We have also investigated the effects of post-selecting on the case in which all of the atoms are found in their excited state instead of the ground state. The value of the fraction $F$ oscillates in a manner similar to that shown in Fig. 5, except that the location of the maxima and minima are interchanged. The physical origin of these effects will be discussed in Section V.

The effects of atomic saturation on the post-selection process can be seen in more detail in the plot of the Q-function in Fig. 6. These results correspond to a coupling parameter of $r = 0.45$, which gives the maximum amount of attenuation. It can be seen that the center of the Q-function has been moved closer to the origin as in Fig. 4. The mean photon number was found to be $\langle \hat{n} \rangle = 2.72$ for normal absorption and $\langle \hat{n} \rangle = 1.04$ for post-selection. In this case, the probability of success is $P_{10} \sim 10^{-4}$, which is lower than before since it is less likely for all the atoms to remain in their ground state when the strength of the interaction is increased.

The Q-function is plotted in Fig. 7 for a value of $r = 0.6$, which corresponds to the maximum amplification. In this case, the mean photon number was found to be $\langle \hat{n} \rangle = 2.39$ and $\langle \hat{n} \rangle = 19.15$ with and without post-selection respectively. This corresponds to nearly a factor of two increase in the photon number as compared to the initial value, or an intensity gain of 2. This post-selection process is not equivalent to a true noiseless amplifier [6] due to the distortion in the shape of the Q-function, but it may still have some advantages over a conventional amplifier as will be discussed below. Surprisingly, the probability of success for the conditions of Fig. 7 is $\sim 10^{-3}$, which is larger than that for the maximum attenuation shown in Fig. 6.

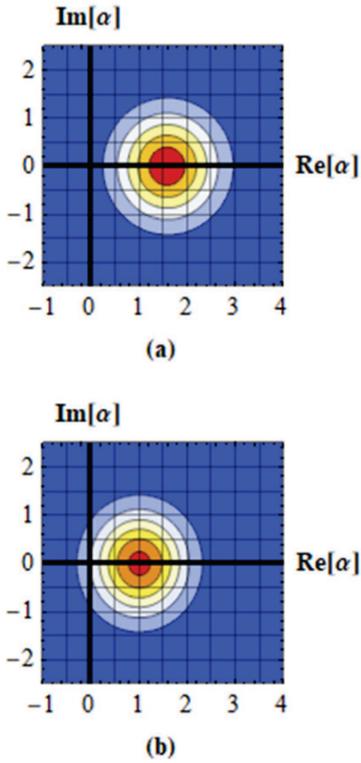

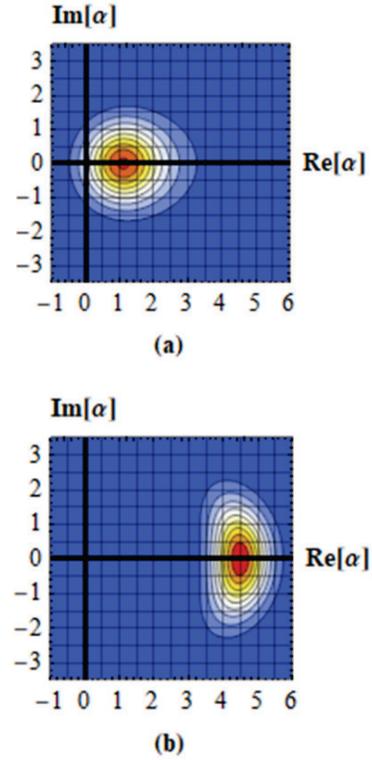

**Figure 6:** The Q-function of the output state for a strong coupling parameter ($r = 0.45$) that gives the maximum attenuation of the signal. (a) Normal absorption, where a small amount of distortion in the quasi-probability distribution is visible. (b) The results of post-selection, which gives more attenuation than is the case for normal absorption. The value of $\alpha$ was once again chosen to be $\sqrt{10}$ and $N = 10$.

**Figure 7:** The Q-function of the output state for a coupling parameter ($r = 0.6$) that gives amplification. (a) Normal attenuation. (b) Results of post-selection, which gives a net amplification for this value of $r$. Substantial distortion of the Q-function can be seen here as in Fig. 6.

Fig. 8 shows the fraction $F$ of photons remaining as a function of the amplitude $|\alpha|$ of the incident coherent state. Here the coupling parameter was held constant at $r = 0.25$. It can be seen that there is no significant difference between the post-selected and ordinary absorption cases in the limit of small $|\alpha|$, as can be shown to be the case analytically. There is also no difference between the two cases in the limit of large $|\alpha|$. This can be understood from the fact that sufficiently large photon numbers will give rapid Rabi oscillations between the two atomic states. This corresponds to saturated absorption where the ground state and excited state are nearly equally populated and post-selection has very little net effect.

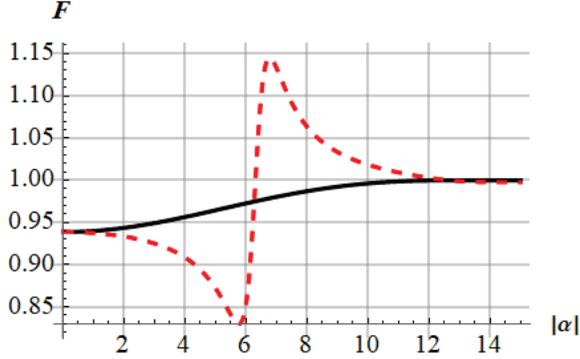

**Figure 8:** A plot of the fraction $F$ of photons remaining as a function of the input amplitude $|\alpha|$ after interaction with a single atom with a coupling strength $r = 0.25$. As before, the black (solid) curve shows normal absorption and the red (dashed) curve shows post-selection. We see that the two curves converge in the limit of both small and large $|\alpha|$.

It is also interesting to investigate the effects of the post-selection process for nonclassical states, such as a squeezed coherent state. This can be calculated using the probability amplitude $c_n$ for a squeezed coherent state in a basis of number (Fock) states, which is given by [20]

$$c_n = A \frac{\left(\frac{1}{2} e^{i\theta} \tanh s\right)^{n/2}}{\sqrt{n! \cosh s}} H_n \left\{ \frac{\gamma}{\sqrt{e^{i\theta} \sinh(2s)}} \right\}. \quad (26)$$

Here $\alpha$ is the amplitude of the initial coherent state while

$$\gamma = \alpha \cosh s + \alpha^* e^{i\theta} \sinh s, \quad (27)$$

$N$ is a normalization constant dependent on $\gamma$, $s$ is the squeezing parameter, and $\theta$ specifies the angle of squeezing.

Fig. 9 shows the Q-function of a squeezed coherent state before and after a post-selection process that produces amplification as in Fig. 8. These results correspond to a squeezing parameter $s = 0.2$ and an amplitude of $\alpha = \sqrt{10}$. It can be seen that amplification still occurs, although the amount of amplification is significantly less than for the coherent state shown in Fig. 8. The amount of squeezing also appears to be increased, which is another indication that the process is nonlinear.

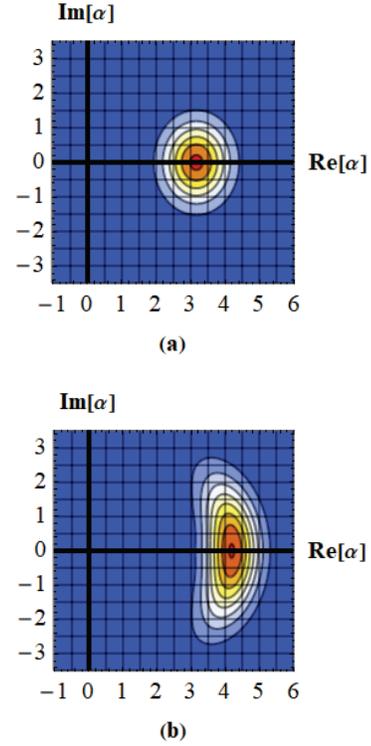

**Figure 9:** A squeezed coherent state amplified using the post-selection process illustrated in Fig. 1. (a) The initial squeezed coherent state with $\alpha = \sqrt{10}$, $\theta = 0$, and $s = 0.2$. (b) The state following post-selection on the atomic ground states for $N = 10$ and $r = 0.6$ as in Fig. 7.

## V. Discussion and Summary

We have shown that post-selection on the ground state of an ensemble of atoms can give either increased attenuation or amplification of an incident coherent state, depending on the strength of the interaction between the atoms and the field. This is a somewhat surprising result, since one might have expected that the atomic medium would have no effect if there are no atomic transitions.

The origin of the increased attenuation for relatively small values of the coupling parameter $r$ is illustrated in Fig. 2. The attenuation is dependent on the fact that the initial number of photons is uncertain.

An event in which no atoms have made a transition to the excited state is less likely to occur for number states $|n\rangle$ with relatively large values of $n$, since the matrix elements for the absorption of a photon are proportional to $\sqrt{n}$. As a result, large values of $n$ are less likely to appear in the post-selected state and the values of $c_n$ are reduced more for large $n$ than they are for smaller values of $n$ as illustrated by the open dots in the figure. This effect reduces the mean number of photons after the state is renormalized.

The origin of the amplification that occurs for certain values of $r$ can be understood from the fact that the atoms will undergo Rabi oscillations when the coupling parameter is sufficiently large. In that case, components with larger values of $n$ may be closer to completing a full Rabi cycle back to the ground state, since the matrix element is larger for large values of $n$. This means that components with large $n$ will now be more likely to give rise to a post-selected state with an atom in the ground state than is the case for smaller values of $n$, which will not have completed a full Rabi oscillation. As a result, the mean photon number will be increased for those values of $r$ rather than decreased. This mechanism explains the resemblance between the oscillatory behavior seen in Fig. 5 and typical plots of Rabi oscillations [17].

The post-selection process does not physically add or remove any photons or energy from the system. Instead, it redistributes the probability amplitudes for the various numbers of photons within the initial uncertainty in $n$. Since this effect is dependent on having an uncertain number of photons, there would be no attenuation or amplification due to post-selection for an incident number state.

Post-selection is generally a nonlinear process, and the results of this approach are not in general equivalent to linear absorption or gain. This can be seen in the distortion of the Q-function in Figs. 6, 7, and 9, as well as the nonlinear dependence on $|\alpha|$ in Fig. 8.

We have not yet discussed the question of how such a post-selection process could be performed experimentally. In principle, an auxiliary field could be used to probe the state of the atoms, but that would involve making separate measurements on each of the atoms. That approach would only be practical for a relatively small number of atoms. A more feasible approach for somewhat larger numbers of atoms would be to use an array of detectors to observe any secondary photons that are subsequently emitted by atoms left in the excited state. Experiments of that kind may be feasible using a nanofiber where the interaction region is relatively small and could be focused on a set of detectors, for example.

The probability of success for the post-selection process decreases exponentially for large coherent state amplitudes. With $\alpha = \sqrt{10}$ the probability of success is typically $\sim 10^{-3}$ for most of the situations considered here. As a result, this technique is limited to relatively weak coherent states in any practical applications.

Amplification of this kind may have some benefits when applied to quantum superposition states, such as Schrodinger cats. If a conventional amplifier, such as an OPA, is used to amplify a superposition state, there will be some amount of quantum noise in the output due to vacuum fluctuations in the input to the idler mode. We recently showed that there will be an additional source of decoherence due to which-path information left in the output idler mode [21], and this can often be much more of a problem than the amplifier noise. The amplification process described here eliminates any which-path information left in the environment and it is therefore capable of amplifying Schrodinger cat states with much less decoherence than an OPA. Post-selecting on an atomic medium may have practical advantages over other forms of noiseless amplification [6-9] due to its relative simplicity, especially for coherent states with relatively large numbers of photons. The distortion seen in Fig. 9 would occur for both components of a Schrodinger cat and it should not reduce the amount of interference between the two components of a cat state [21] as a result.

In summary, we have shown that increased attenuation of a coherent state by an atomic medium can occur in the absence of any actual absorption. This counter-intuitive result is somewhat similar to our earlier work on noiseless attenuation using an OPA [12]. Optical amplification instead of attenuation can also occur for certain values of the coupling parameter $r$. These results are of fundamental interest and they represent a new method for optical amplification. This approach may be of practical use in quantum communications or quantum sensor systems that utilize macroscopic quantum superposition states.

**Acknowledgements**

This work was supported in part by the National Science Foundation under grant number PHY-1802472.